\documentclass[11pt]{article}%
\usepackage{amsmath}
\usepackage{graphicx}%
\usepackage{amsfonts}%
\usepackage{amssymb}
\newcommand{\eqnb}{\begin{equation}}
\newcommand{\eqne}{\end{equation}}

\newtheorem{The}{Theorem}

\begin{document}

\title{A Unified Framework for Analyzing Closed Queueing Networks in Bike Sharing Systems\footnote{{Interpretation:} After this paper was published, we find an interesting work by Rick Zhang and Marco Pavone ``R. Zhang and M. Pavone (2014). A queueing network approach to the analysis and control of mobility-on-demand systems. Published Online: arXiv:1409.6775, Pages 1-9."}}
\author{Quan-Lin Li, Rui-Na Fan and Jing-Yu Ma\\School of Economics and Management Sciences \\Yanshan University, Qinhuangdao 066004, P.R. China}
\date{Published in {\bf Information Technologies and Mathematical Modelling}, Springer, 2016}
\maketitle

\begin{abstract}
During the last decade bike sharing systems have emerged as a public transport
mode in urban short trips in more than 500 major cities around the world. For
the mobility service mode, many challenges from its operations are not well
addressed yet, for example, how to develop the bike sharing systems to be able
to effectively satisfy the fluctuating demands both for bikes and for vacant
lockers. To this end, it is a key to give performance analysis of the bike
sharing systems. This paper first describes a large-scale bike sharing system.
Then the bike sharing system is abstracted as a closed queueing network with
multi-class customers, where the virtual customers and the virtual nodes are
set up, and the service rates as well as the relative arrival rates are
established. Finally, this paper gives a product-form solution to the steady
state joint probabilities of queue lengths, and gives performance analysis of
the bike sharing system. Therefore, this paper provides a unified framework
for analyzing closed queueing networks in the study of bike sharing systems.
We hope the methodology and results of this paper can be applicable in the
study of more general bike sharing systems.

\vskip                                                              0.5cm


\end{abstract}

\section{Introduction}
During the last decade the bike sharing systems are fast increasing as a
public transport mode in urban short trips, and have been launched in more
than 500 major cities around the world. Also, the bike sharing systems offer a
low cost and environmental protection\ mobility service through sharing
one-way use. Now, the bike sharing systems are regarded as an effective way to
jointly solve traffic congestion, parking difficulties, traffic noise, air
pollution and so forth. DeMaio \cite{DeM:2009} reviewed the history, impacts,
models of provision and future of the bike sharing systems. Larsen
\cite{Lar:2013} reported that over 500 major cities host advanced bike sharing
systems with a combined fleet of more than half a million bikes up to April
2013. A synthesis of the literature for the bike sharing systems was given by
Fishman et al. \cite{Fis:2013} and Labadi et al. \cite{Lab:2015}. At the same
time, for some countries or cities developing the bike sharing systems,
readers may refer to, such as, Europe, the Americas and Asia by Shaheen et al.
\cite{Sha:2010}, the European OBIS Project by Janett and Hendrik
\cite{Jan:2011}, the France by Faye \cite{Fay:2008}, China by Tang et al.
\cite{Tang:2011}, London by Lathia et al. \cite{Lat:2012}, Montreal by Morency
et al. \cite{Mor:2011}, and a number of famous cities by Shu et al.
\cite{Shu:2013}.

In operations of the bike sharing systems, a crucial question is the ability
not only to meet the fluctuating demand for renting bikes at each station but
also to provide enough vacant lockers to allow the renters to return the bikes
at their destinations. Since the number of bikes packed in each station is
always randomly dynamically changed, this causes an unpredictable imbalance,
such as, some stations contain more bikes but the others are seriously short
of available bikes. Such a randomly dynamic unbalance of bikes distributed
among the stations often leads to occurrence of the problematic stations
(i.e., full or empty stations). Notice that the problematic stations reflect a
common challenge faced by the bike sharing systems in practice due to the
stochastic and time-inhomogeneous nature of both the customer arrivals and the
bike returns, thus the probability of problematic stations has been regarded
as a main factor to measure the satisfaction of customers and even to estimate
the quality of service. Obviously, how to effectively reduce the probability
of problematic stations becomes a key way to improve the satisfaction of
customers and further to promote the quality of system service. Therefore, it
is a major task to develop effective methods for computing the probability of
problematic stations in the study of bike sharing systems.

Queueing theory and Markov processes are very useful for computing the
probability of problematic stations, and more generally, analyzing performance
measures of the bike sharing systems. However, available works on such a
research line are still fewer up to now. We would like to refer readers to
four classes of recent literature as follows. \textbf{(a) Simple queues:}
Leurent \cite{Leu:2012} used the $M/M/1/C$ queue to study a vehicle-sharing
system in which each station contains an additional waiting room which helps
those customers arriving at a problematic station, and analyzed performance
measures of this system in terms of a geometric distribution. Schuijbroek et
al. \cite{Sch:2013} evaluated the service level by means of the transient
distribution of the $M/M/1/C$ queue, and the service level is used to
establish some optimal models to discuss the inventory rebalancing and vehicle
routing. Raviv et al \cite{Rav:2013} and Raviv and Kolka \cite{Rav:2013a}
employed the transient distribution of a time-inhomogeneous $M\left(
t\right)  /M\left(  t\right)  /1/C$ queue to compute the expected number of
bike shortages at each station. \textbf{(b) The mean-field theory:} Fricker et
al. \cite{Fri:2012} considered a space inhomogeneous bike sharing system with
different clusters, and expressed the minimal proportion of problematic
stations within each cluster. For a space homogeneous bike sharing system,
Fricker and Gast \cite{Fri:2014} used the $M/M/1/K$ queue to provide a more
detailed analysis for some simple mean-field models (including the
\textit{power of two choices}), derived a closed-form solution to the minimal
proportion of problematic stations, and compared the incentives and
redistribution mechanisms. Fricker and Tibi \cite{Fri:2015} studied the
central limit and local limit theorems for the independent (perhaps non
identically distributed) random variables which effectively support analysis
of a generalized Jackson network with product-form solution; and used these
obtained results to evaluate performance measures of the space inhomogeneous
bike sharing systems, where its asymptotics gives a complete picture for
equilibrium state analysis of the locally space homogeneous bike sharing
systems. Li et al. \cite{Li:2016} provided a mean-field queueing method to
study a large-scale bike sharing system through using a combination of, such
as, the virtual time-inhomogeneous queue, the mean-field equations, the
martingale limit, the nonlinear birth-death process, numerical computation of
the fixed point, and numerical analysis for the steady state probability of
the problematic stations. \textbf{(c) Queueing networks:} Savin et al.
\cite{Sav:2005} used a loss network as well as admission control to discuss
capacity allocation of a rental model with two classes of customers, and
studied the revenue management and fleet sizing decision in the rental system.
Adelman \cite{Ade:2007} applied a closed queueing network to set up an
internal pricing mechanism for managing a fleet of service units, and also
used a nonlinear flow model to discuss the price-based policy for the vehicle
redistribution. George and Xia \cite{Geo:2011} provided a queueing network
method in the study of vehicle rental systems, and determined the optimal
number of parking spaces for each rental location. \textbf{(d) Markov decision
processes:} Stochastic optimization and Markov decision processes are applied
to analysis of the bike sharing systems. From a dynamic price mechanism,
Waserhole and Jost \cite{Was:2012} used the closed queuing networks to propose
a Markov decision model of a bike sharing system. To overcome the curse of
dimensionality in the Markov decision process with a high dimension, they
established a fluid approximation that computes a static policy and gave an
upper bound on the potential optimization. Such a fluid approximation for the
Markov decision processes of the bike sharing systems was further developed in
Waserhole and Jost \cite{Was:2013} \cite{Was:2014} and Waserhole et al.
\cite{Was:2013a}.

The main purposes of this paper are to provide a unified framework for
analyzing closed queueing networks in the study of bike sharing systems. This
framework of closed queueing networks is interesting, difficult and
challenging from three crucial features: (a) Stations and roads have very
different physical attributes, but all of them are abstracted as
indistinguishable nodes in the closed queueing networks; (b) the service
discipline of the stations is First Come First Service (abbreviated as FCFS),
while the service discipline of the roads is Processor Sharing (abbreviated as
PS); and (c) the virtual customers (i.e., bikes) in the stations are of a
single class, while the virtual customers (i.e., bikes) in the roads are of
two classes, and their classes may change on the roads according to the first
bike-return or the at least two successive bike-returns due to the full
stations, respectively. For such a closed queueing network, this paper
provides a detailed analysis both for establishing a product-form solution to
the steady state joint probabilities of queue lengths, and for computing the
steady state probability of problematic stations, more generally, for
analyzing performance measures of the bike sharing system. The main
contributions of this paper are twofold. The first contribution is to describe
a large-scale bike sharing system and to provide a unified framework for
analyzing closed queueing networks through establishing some basic factors:
The service rates from stations or roads; and the routing matrix as well as
the relative arrival rates to stations or roads. Notice that the basic factors
play a key role in the study of closed queueing networks. The second
contribution of this paper is to provide a product-form solution to the
steady state joint probabilities of queue lengths in the closed queueing
network, and give performance analysis of the bike sharing system in terms of
the steady state joint probabilities.

The remainder of this paper is organized as follows. In Section 2, we describe
a large-scale bike sharing system with $N$ different stations and with at most
$N\left(  N-1\right)  $ different roads. In Section 3, we provide a unified
framework for analyzing closed queueing networks in the study of bike sharing
systems, and also compute the service rates, the routing matrix, and the
relative arrival rates. In Section 4, we give a product-form solution to the
steady state joint probabilities of queue lengths in the closed queueing
network, and analyze performance measures of the bike sharing system by means
of the steady state joint probabilities. Some concluding remarks are given in
Section 5.

\section{Model Description}

In this section, we describe a large-scale bike sharing system with $N$
different stations and with at most $N\left(  N-1\right)  $ different roads
due to the riding-bike directed connection between any two stations. To
analyze such a bike sharing system, we provide a unified framework for
analyzing closed queueing networks in the study of bike sharing systems.

In a large-scale bike sharing system, a customer arrives at a station, rents a
bike, and uses it for a while; then she returns the bike to a destination
station, and immediately leaves this system. Obviously, for any customer
renting and using a bike, her first return-bike time is different from those
return-bike times that she has successively returned the bike for at least two
times due to arriving at the full stations. At the same time, it is easy to
understand that for any customer, her first road selection as well as her
first riding-bike speed are different from those of having successively
returned her bike for at least two times. Also, it is noted that the customer
must return her bike to a station, then she can immediately leave the bike
sharing system.

Now, we describe the bike sharing system, including operation mechanism,
system parameters and mathematical notation, as follows:

\textbf{(1) Stations and roads: }There are $N$ different stations and at most
$N\left(  N-1\right)  $ different roads, where the $N\left(  N-1\right)  $
roads are observed from the fact that there must exist a directed road from a
station to another station. In addition, we assume that at the initial time
$t=0$, every station has $C$ bikes and $K$ parking places, where $1\leq
C<K<\infty$; and $NC\geq K$, which makes that some of the $NC$ bikes can
result in at least a full station.

\textbf{(2) Customer arrival process: }The arrivals of the outside customers
at the $i$th station are a Poisson process with arrival rate $\lambda_{i}>0$
for $1\leq i\leq N$.

\textbf{(3) The first riding-bike time:} Once an outside customer arrives at
the $i$th station, she immediately goes to rent a bike. If there is no bike in
the $i$th station (i.e., the $i$th station is empty), then the customer
directly leaves this bike sharing system. If there is at least one bike in the
$i$th station, then the customer rents a bike, and then goes to Road
$i\rightarrow j$. We assume that for $j\neq i$ with $1\leq i,j\leq N$, the
customer at the $i$th station rides the bike into Road $i\rightarrow j$ with
probability $p_{i,j}$ for $\sum_{j\neq i}^{N}p_{i,j}=1$; and her riding-bike
time from the $i$th station to the $j$th station (i.e., riding on Road
$i\rightarrow j$) is an exponential random variable with riding-bike rate
$\mu_{i,j}>0$, where the expected riding-bike time is $1/\mu_{i,j}$.

\textbf{(4) The bike return times:}

\underline{The first return} -- When the customer completes her short trip on
the above Road $i\rightarrow j$ (see Assumption (3)), she needs to return her
bike to the $j$th station. If there is at least one available parking position
(i.e., a vacant docker), then the customer directly returns her bike to the
$j$th station, and immediately leaves this bike sharing system.

\underline{The second return} -- If no parking position is available at the
$j$th station, then she has to ride the bike to another station $l_{1}$ with
probability $\alpha_{j,l_{1}}$ for $l_{1}\neq j$ for $\sum_{l_{1}\neq j}%
^{N}\alpha_{j,l_{1}}=1$; and her riding-bike time from the $j$th station to
the $l_{1}$th station (i.e., riding on Road $j\rightarrow l_{1}$) is an
exponential random variable with riding-bike rate $\xi_{j,l_{1}}>0$. If there
is at least one available parking position, then the customer directly returns
her bike to the $l_{1}$th station, and immediately leaves this bike sharing system.

\underline{The third return} -- If no parking position is available at the
$l_{1}$th station, then she has to ride the bike to another station $l_{2}$
with probability $\alpha_{l_{1},l_{2}}$ for $l_{2}\neq l_{1}$ for $\sum
_{l_{2}\neq l_{1}}^{N}\alpha_{l_{1},l_{2}}=1$; and her riding-bike time from
the $l_{1}$th station to the $l_{2}$th station (i.e., riding on Road
$l_{1}\rightarrow l_{2}$) is an exponential random variable with riding-bike
rate $\xi_{l_{1},l_{2}}>0$. If there is at least one available parking
position, then the customer directly returns her bike to the $l_{2}$th
station, and immediately leaves this bike sharing system.

\underline{The ($k+1$)st return for $k\geq3$ }-- We assume that this
bike has not been returned at any station yet through $k$ consecutive return
processes. In this case, the customer has to try her ($k+1$)st lucky return.
Notice that the customer goes to the $l_{k}$th station from the $l_{k-1}$th
full station with probability $\alpha_{l_{k-1},l_{k}}$ for $l_{k}\neq l_{k-1}$
for $\sum_{l_{k}\neq l_{k-1}}^{N}\alpha_{l_{k-1},l_{k}}=1$; and her
riding-bike time from the $l_{k-1}$th station to the $l_{k}$th station (i.e.,
riding on Road $l_{k-1}\rightarrow l_{k}$) is an exponential random variable
with riding-bike rate $\xi_{l_{k-1},l_{k}}>0$. If there is at least one
available parking position, then the customer directly returns her bike to the
$l_{k}$th station, and immediately leaves this bike sharing system; otherwise
she has to continuously try another station again.

We further assume that the returning-bike process is persistent in the sense
that the customer must find a station with an empty position to return her
bike, because the bike is the public property so that no one can make it her own.

It is seen from the above description that the parameters: $p_{i,j}$ and
$\mu_{i,j}$ for $j\neq i$ and $1\leq i,j\leq N$, of the first return, may be
different from the parameters: $\alpha_{i,j}$ and $\xi_{i,j}$ for $j\neq i$
and $1\leq i,j\leq N$, of the $k$th return for $k\geq2$. Notice that such an
assumption with respect to these different parameters is actually reasonable
because the customer possibly has more things (for example, tourism, shopping,
visiting friends and so on) in the first return process, but she become to
have only one return task during the $k$ successive return processes for
$k\geq2$.

\textbf{(5) The departure discipline:} The customer departure has two
different cases: (a) An outside customer directly leaves the bike sharing
system if she arrives at an empty station; or (b) if one customer rents and
uses a bike, and she finally returns the bike to a station, then the customer
completes her trip, and immediately leaves the bike sharing system.

We assume that the customer arrival and riding-bike processes are independent,
and also all the above random variables are independent of each other. For
such a bike sharing system, Figure 1 provides some physical interpretation.

\begin{figure}[ptb]
\setlength{\abovecaptionskip}{0.cm}  \setlength{\belowcaptionskip}{-0.cm}
\centering               \includegraphics[width=9cm]{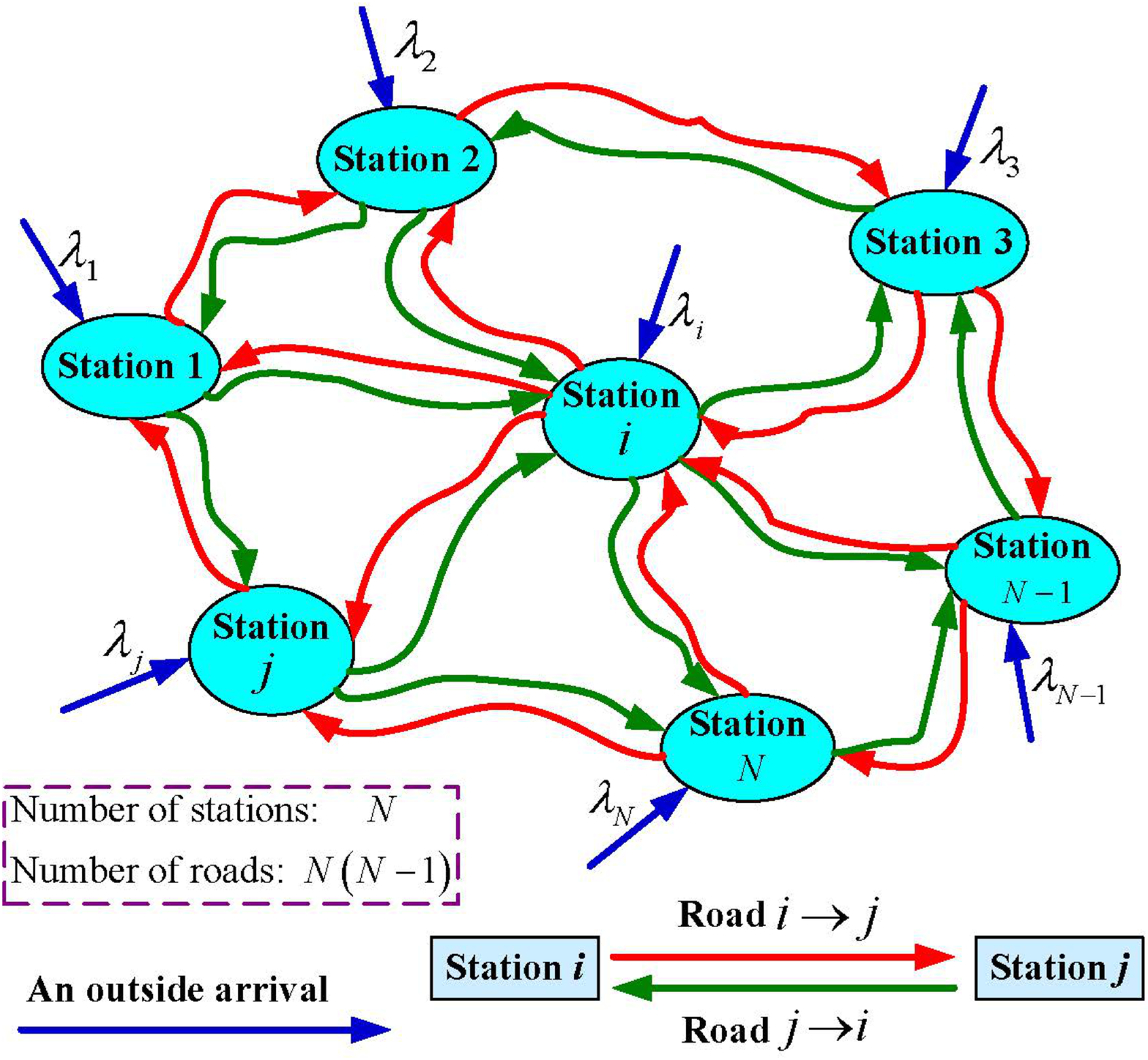}
\newline \caption{The physical structure of the bike sharing system}%
\label{figure:fig-1}%
\end{figure}

\section{A Closed Queueing Network}

In this section, we first provide a closed queueing network to express the
bike sharing system, as seen in Figure 1. Then we determine the service rates,
the routing matrix, and the relative arrival rates of the closed queueing
network. Notice that there are two classes of customers in the $N\left(
N-1\right)  $ roads.

In the bike sharing system described in the above section, there are $NC$
bikes, $N$ stations and $N\left(  N-1\right)  $ roads. Now, we abstract the
bike sharing system as a closed queueing network as follows:

\begin{description}
\item[\textbf{(1) Virtual nodes:} ] Although the stations and roads have
different physical attributes such as functions, and geographical topologies,
the stations and roads are all regarded as the same nodes in the closed
queueing network.

\item[\textbf{(2) Virtual customers:}] The bikes at the stations or roads are
described as follows:
\end{description}

\underline{Abstract:} The virtual customers are abstracted by the bikes,
which are either parked in the stations or ridden on the roads. Notice that
the total number of bikes in the bike sharing system is fixed as $NC$ due to
the fact that bikes can neither enter nor leave this system, thus the bike
sharing system can be regarded as a closed queueing network.

\underline{Multiple classes:} From Assumption (2) in Section 2, it is seen
that there are only one class of customers in the nodes abstracted from the
stations. From Assumptions (3) and (4) in Section 2, we understand that there
are two different classes of customers in the nodes abstracted from the roads,
where the first class of customers are the bikes ridden on the roads for the
first time; while the second class of customers are the bikes which are
successively ridden on the at least two different roads due to the full station.

\begin{description}
\item[(3) Service disciplines:] The First Come First Service (or FCFS) is used
in the nodes abstracted from the stations; while a new processor sharing (or
PS) is used in the nodes abstracted from the roads.
\end{description}

In the above closed queueing network, let $Q_{i}\left(  t\right)  $\ be the
number of bikes parked in $i$th station at time $t\geq0$ for $1\leq i\leq N$,
and $R_{k,l}^{\left(  r\right)  }\left(  t\right)  $ the number of bikes of
class $r$ ridden on Road $k\rightarrow l$ at time $t$ for $r=1,2$, and $k\neq
l$ with $1\leq k,l\leq N$. We write%
\[
\mathbf{X}\left(  t\right)  =\left(  \mathbf{L}_{1}\left(  t\right)
,\mathbf{L}_{2}\left(  t\right)  ,\ldots,\mathbf{L}_{N-1}\left(  t\right)
,\mathbf{L}_{N}\left(  t\right)  \right)  ,
\]
where for $1\leq i\leq N$%
\begin{align*}
\mathbf{L}_{i}\left(  t\right)  = &  \left(  Q_{i}\left(  t\right)
;R_{i,1}^{\left(  1\right)  }\left(  t\right)  ,R_{i,1}^{\left(  2\right)
}\left(  t\right)  ;R_{i,2}^{\left(  1\right)  }\left(  t\right)
,R_{i,2}^{\left(  2\right)  }\left(  t\right)  ;\ldots;R_{i,i-1}^{\left(
1\right)  }\left(  t\right)  ,R_{i,i-1}^{\left(  2\right)  }\left(  t\right)
;\right.  \\
&  \left.  R_{i,i+1}^{\left(  1\right)  }\left(  t\right)  ,R_{i,i+1}^{\left(
2\right)  }\left(  t\right)  ;R_{i,i+2}^{\left(  1\right)  }\left(  t\right)
,R_{i,i+2}^{\left(  2\right)  }\left(  t\right)  ;\ldots;R_{i,N}^{\left(
1\right)  }\left(  t\right)  ,R_{i,N}^{\left(  2\right)  }\left(  t\right)
\right)  .
\end{align*}
Obviously, $\left\{  \mathbf{X}\left(  t\right)  :t\geq0\right\}  $ is a
Markov process of size $N\left(  2N-1\right)  $ due to the exponential and
Poisson assumptions of this bike sharing system.

Now, we describe the state space of the Markov process $\left\{
\mathbf{X}\left(  t\right)  :t\geq0\right\}  $. It is seen from Section 2 that%
\begin{equation}
0\leq Q_{i}\left(  t\right)  \leq K,\text{ \ }1\leq i\leq N,\label{Cequ-1}%
\end{equation}%
\begin{equation}
0\leq R_{k,l}^{\left(  r\right)  }\left(  t\right)  \leq NC,\text{
\ \ \ \ }r=1,2,\text{ }k\neq l,\text{ }1\leq k,l\leq N,\label{Cequ-2}%
\end{equation}
and%
\begin{equation}
\sum_{i=1}^{N}Q_{i}\left(  t\right)  +\sum_{k=1}^{N}\sum_{l\neq k}^{N}%
R_{k,l}^{\left(  1\right)  }\left(  t\right)  +\sum_{k=1}^{N}\sum_{l\neq
k}^{N}R_{k,l}^{\left(  2\right)  }\left(  t\right)  =NC.\label{Cequ-3}%
\end{equation}
From (\ref{Cequ-1}) to (\ref{Cequ-3}), it is seen the state space of Markov
process $\left\{  \mathbf{X}\left(  t\right)  :t\geq0\right\}  $ of size
$N\left(  2N-1\right)  $ is given by%
\begin{align*}
\Omega= &  \left\{  \overrightarrow{n}:0\leq n_{i}\leq K,0\leq m_{k,l}%
^{\left(  1\right)  },m_{k,l}^{\left(  2\right)  }\leq NC,\right.  \\
&  \left.  \sum\limits_{i=1}^{N}n_{i}+\sum_{k=1}^{N}\sum_{l\neq k}^{N}%
m_{k,l}^{\left(  1\right)  }+\sum_{k=1}^{N}\sum_{l\neq k}^{N}m_{k,l}^{\left(
2\right)  }=NC\right\}  ,
\end{align*}
where%
\[
\overrightarrow{n}=\left(  \mathbf{n}_{1},\mathbf{n}_{2},\ldots,\mathbf{n}%
_{N-1},\mathbf{n}_{N}\right)  ,
\]
and for $1\leq i\leq N$%
\begin{align*}
\mathbf{n}_{i}= &  \left(  n_{i};m_{i,1}^{\left(  1\right)  },m_{i,1}^{\left(
2\right)  };m_{i,2}^{\left(  1\right)  },m_{i,2}^{\left(  2\right)  }%
;\ldots;m_{i,i-1}^{\left(  1\right)  },m_{i,i-1}^{\left(  2\right)  };\right.
\\
&  \left.  m_{i,i+1}^{\left(  1\right)  },m_{i,i+1}^{\left(  2\right)
};m_{i,i+2}^{\left(  1\right)  },m_{i,i+2}^{\left(  2\right)  };\ldots
;m_{i,N}^{\left(  1\right)  },m_{i,N}^{\left(  2\right)  }\right)  .
\end{align*}
Notice that $m_{k,l}=m_{k,l}^{\left(  1\right)  }+m_{k,l}^{\left(  2\right)
}$ is the total number of bikes being ridden on Road $k\rightarrow l$ for
$k\neq l$ with $1\leq k,l\leq N$, and also the state space $\Omega$ contains
$\left(  K+1\right)  ^{N}\left(  NC+1\right)  ^{2N\left(  N-1\right)  }$ elements.

To compute the steady state joint probabilities of $N\left(  2N-1\right)  $
queue lengths in the bike sharing system, it is seen from Chapter 7 in Bolch
et al. \cite{Bol:2006} that we need to determine the service rate, the routing
matrix and the relative arrival rate for each node in the closed queueing network.

\textbf{(a) The service rates}

From Figure 2, it is seen that the service rates of the closed queueing
network are given from two different cases as follows:

\begin{figure}[ptb]
\setlength{\abovecaptionskip}{0.cm}  \setlength{\belowcaptionskip}{-0.cm}
\centering                \includegraphics[width=9cm]{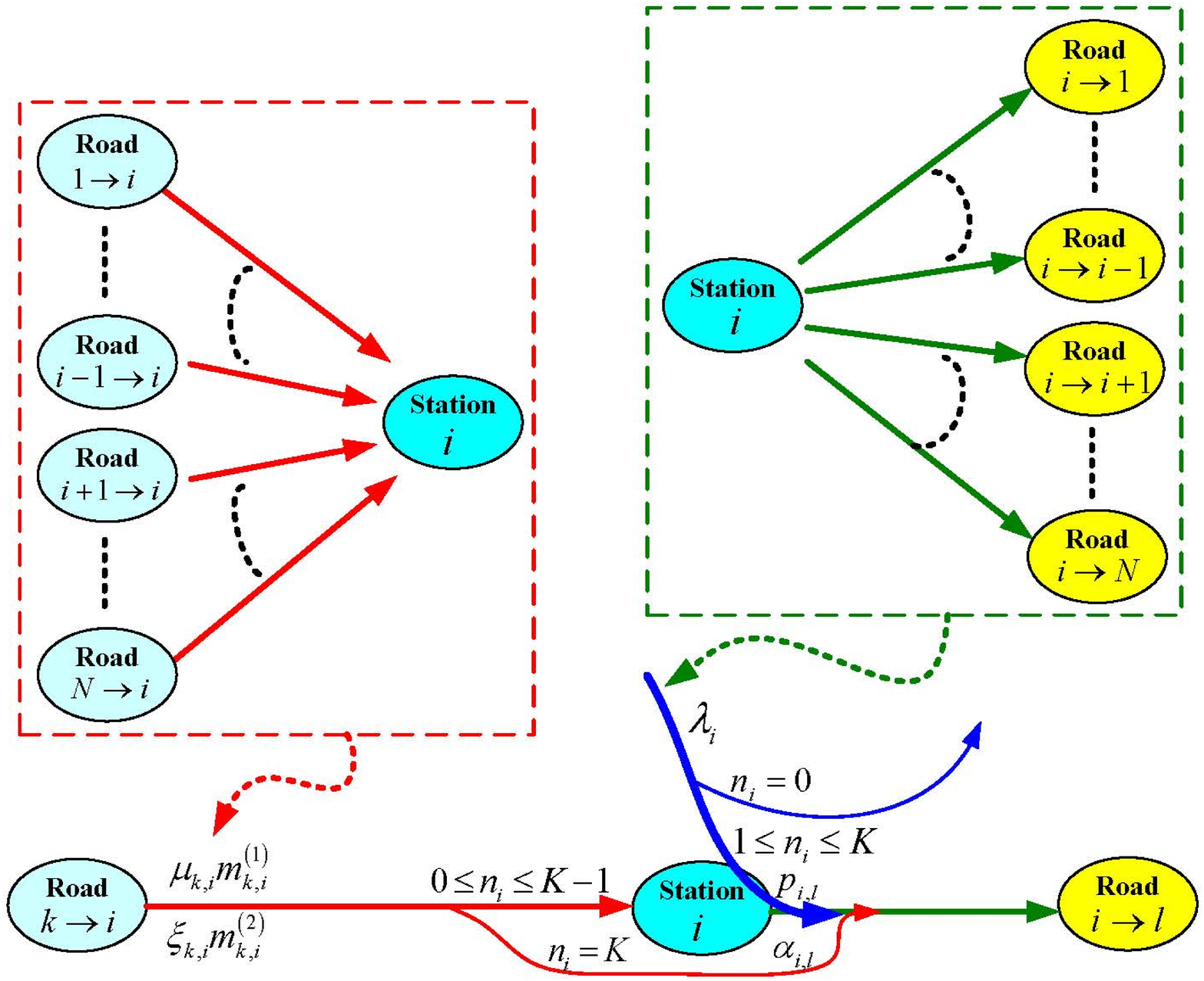}
\newline \caption{The queueing processes in the closed queueing network}%
\label{figure:fig-2}%
\end{figure}

\textit{Case one: The node is one of the }$N$\textit{ stations}

The departure process of bikes from the $i$th station, renting at the $i$th
station and immediately ridden on one of the $N-1$ roads (such as, Road
$i\rightarrow l$ for $l\neq i$ with $1\leq l\leq N$), is Poisson with service
rate%
\begin{equation}
b_{i}=\lambda_{i}\mathbf{1}_{\left\{  1\leq n_{i}\leq K\right\}  }\sum_{l\neq
i}^{N}p_{i,l}=\lambda_{i}\mathbf{1}_{\left\{  1\leq n_{i}\leq K\right\}
}\label{Cequ-5}%
\end{equation}
by means of the condition: $\sum_{l\neq i}^{N}p_{i,l}=1$.

\textit{Case two: The node is one of the }$N\left(  N-1\right)  $\textit{ roads}

In this case, two different processor sharing queueing processes of Road
$i\rightarrow l$ (with two classes of different customers) are explained in
Figure 2. Now, we describe the service rates with respect to the two classes
of different customers as follows:

\underline{The first class of customers:} The departure process of bikes from
Road $i\rightarrow l$, rented from Station $i$ and being ridden on Road
$i\rightarrow l$ for the first time, is Poisson with service rate%
\begin{equation}
b_{i,l}^{\left(  1\right)  }=m_{i,l}^{\left(  1\right)  }\mu_{i,l}.
\label{Cequ-6}%
\end{equation}

\underline{The second class of customers:} The departure process of bikes from
Road $i\rightarrow l$, having successively been ridden on the roads for at least
two times but now on Road $i\rightarrow l$, is Poisson with service rate%
\begin{equation}
b_{i,l}^{\left(  2\right)  }=m_{i,l}^{\left(  2\right)  }\xi_{i,l}.
\label{Cequ-7}%
\end{equation}

\textbf{(b) The routing matrix and the relative arrival rates}

Now, we compute the relative arrival rate of each node in the closed queueing
network. Differently from the service rates analyzed above, it is more
complicated to determine the relative arrival rates by means of the routing matrix.

Based on Chapter 7 in Bolch et al. \cite{Bol:2006}, we denote by $e_{i}\left(
n_{i}\right)  $ and $e_{i,l}^{\left(  r\right)  }\left(  m_{i,l}^{\left(
r\right)  }\right)  $ the relative arrival rates of the $i$th station with
$n_{i}$ parking bikes, and of Road $i\rightarrow l$ with $m_{i,l}^{\left(
r\right)  }$ riding bikes of class $r$,\ respectively. We write%
\[
\mathbb{E}=\left\{  \overrightarrow{e}\left(  \overrightarrow{n}\right)
:\overrightarrow{n}\in\Omega\right\}  ,
\]
where%
\[
\overrightarrow{e}\left(  \overrightarrow{n}\right)  =\left(  \mathbf{e}%
_{1}\left(  \overrightarrow{n}\right)  ,\mathbf{e}_{2}\left(  \overrightarrow
{n}\right)  ,\ldots,\mathbf{e}_{N-1}\left(  \overrightarrow{n}\right)
,\mathbf{e}_{N}\left(  \overrightarrow{n}\right)  \right)  ,
\]
and for $1\leq i\leq N$%
\begin{align*}
\mathbf{e}_{i}\left(  \overrightarrow{n}\right)  = &  \left(  e_{i}\left(
n_{i}\right)  ;e_{i,1}^{\left(  1\right)  }\left(  m_{i,1}^{\left(  1\right)
}\right)  ,e_{i,1}^{\left(  2\right)  }\left(  m_{i,1}^{\left(  2\right)
}\right)  ;\ldots;e_{i,i-1}^{\left(  1\right)  }\left(  m_{i,i-1}^{\left(
1\right)  }\right)  ,e_{i,i-1}^{\left(  2\right)  }\left(  m_{i,i-1}^{\left(
2\right)  }\right)  ;\right.  \\
&  \left.  e_{i,i+1}^{\left(  1\right)  }\left(  m_{i,i+1}^{\left(  1\right)
}\right)  ,e_{i,i+1}^{\left(  2\right)  }\left(  m_{i,i+1}^{\left(  2\right)
}\right)  ;\ldots;e_{i,N}^{\left(  1\right)  }\left(  m_{i,N}^{\left(
1\right)  }\right)  ,e_{i,N}^{\left(  2\right)  }\left(  m_{i,N}^{\left(
2\right)  }\right)  \right)  .
\end{align*}

Now, we introduce two useful notations: $\overrightarrow{g}_{i}$ and
$\overrightarrow{g}_{i,l}^{\left(  r\right)  }$ as follows:

\begin{description}
\item[$\overrightarrow{g}_{i}:$] A unit row vector of size $N\left(  2N-1\right)  $, which is given by a method of
replacing elements from $\overrightarrow{n}$ to $\overrightarrow{g}_{i}$, that
is, corresponding to the row vector $\overrightarrow{n}$, the element $n_{i}$
is replaced by one, while all other elements of the vector $\overrightarrow
{n}$ are replaced by zeros.

\item[$\overrightarrow{g}_{i,l}^{\left(  r\right)  }:$] A unit row vector of
size $N\left(  2N-1\right)  $, which is
given by a method of replacing elements from $\overrightarrow{n}$ to
$\overrightarrow{g}_{i,l}^{\left(  r\right)  }$, that is, corresponding to the
row vector $\overrightarrow{n}$, the element $m_{i,l}^{\left(  r\right)  }$ is
replaced by one, while all other elements of the vector $\overrightarrow{n}$
are replaced by zeros.
\end{description}

To compute the vector $\overrightarrow{e}\left(  \overrightarrow{n}\right)  $,
we first need to give the routing matrix $\mathbf{P}$ of the closed queueing
network as follows:%
\[
\mathbf{P}=\left(  P_{\overrightarrow{n},\overrightarrow{n}^{\prime}}\right)
_{\overrightarrow{n},\overrightarrow{n}^{\prime}\in\Omega},
\]
where the routing matrix $\mathbf{P}$ is of order $\left(  K+1\right)  ^{N}\left(  NC+1\right)  ^{2N\left(  N-1\right)  }$, and the element
$P_{\overrightarrow{n},\overrightarrow{n}^{\prime}}$ is computed from the
following three cases:

\textit{Case one: From a station to a road}

For $1\leq i,l\leq N$ with $l\neq i$, we observe a transition route from the
$i$th station to Road $i\rightarrow l$. If a rented bike leaves the $i$th
station and enters Road $i\rightarrow l$, then $1\leq n_{i}\leq K$, and there
is a two-element change: $\left(  n_{i},m_{i,l}^{\left(  1\right)  }\right)
\rightarrow\left(  n_{i}-1,m_{i,l}^{\left(  1\right)  }+1\right)  $. Thus we
obtain that for $1\leq n_{i}\leq K$%
\[
P_{\overrightarrow{n},\overrightarrow{n}^{\prime}}=P_{\overrightarrow
{n},\overrightarrow{n}-\overrightarrow{g}_{i}+\overrightarrow{g}%
_{i,l}^{\left(  1\right)  }}=p_{i,l}%
\]
by means of Assumption (3) of Section 2. There are $NK\left(  N-1\right)  $
such elements with $P_{\overrightarrow{n},\overrightarrow{n}^{\prime}%
}=P_{\overrightarrow{n},\overrightarrow{n}-\overrightarrow{g}_{i}%
+\overrightarrow{g}_{i,l}^{\left(  1\right)  }}=p_{i,l}$ in the closed
queueing network.

\textit{Case two: From a road to a station}

For $r=1,2$ and $1\leq k,i,l\leq N$ with $i\neq k$ and $l\neq i$, we observe a
transition route from Road $k\rightarrow i$ to the $i$th station. If a riding
bike of class $r$ leaves Road $k\rightarrow i$, then either it enters the
$i$th station if $0\leq n_{i}\leq K-1$; or it goes to Road $i\rightarrow l$ if
$n_{i}=K$.

In the former case (the riding bike of class $r$ enters the $i$th station if
$0\leq n_{i}\leq K-1$), we obtain that for $0\leq n_{i}\leq K-1$, there is a
two-element change: $\left(  m_{k,i}^{\left(  r\right)  },n_{i}\right)
\rightarrow\left(  m_{k,i}^{\left(  r\right)  }-1,n_{i}+1\right)  $, hence
this gives that for $0\leq n_{i}\leq K-1$%
\[
P_{\overrightarrow{n},\overrightarrow{n}^{\prime}}=P_{\overrightarrow
{n},\overrightarrow{n}-\overrightarrow{g}_{k,i}^{\left(  r\right)
}+\overrightarrow{g}_{i}}=1,
\]
since the end of Road $k\rightarrow i$ is only the $i$th station. There are
$2N^{2}\left(  N-1\right)  CK$ such elements with $P_{\overrightarrow
{n},\overrightarrow{n}^{\prime}}=P_{\overrightarrow{n},\overrightarrow
{n}-\overrightarrow{g}_{k,i}^{\left(  r\right)  }+\overrightarrow{g}_{i}}=1$
in the closed queueing network.

\textit{Case three: From a road to another road}

In the latter case (the riding bike of class $r$ goes to Road $i\rightarrow l$
if $n_{i}=K$), we get that there is a two-element change: $\left(
m_{k,i}^{\left(  r\right)  },m_{i,l}^{\left(  2\right)  }\right)
\rightarrow\left(  m_{k,i}^{\left(  r\right)  }-1,m_{i,l}^{\left(  2\right)
}+1\right)  $. Thus we obtain that for $n_{i}=K$%
\[
P_{\overrightarrow{n},\overrightarrow{n}^{\prime}}=P_{\overrightarrow
{n},\overrightarrow{n}-\overrightarrow{g}_{k,i}^{\left(  r\right)
}+\overrightarrow{g}_{i,l}^{\left(  2\right)  }}=\alpha_{i,l}%
\]
by means of Assumption (4) of Section 2. There are $2N^{3}\left(  N-1\right)
^{2}C^{2}$ such elements with $P_{\overrightarrow{n},\overrightarrow
{n}^{\prime}}=P_{\overrightarrow{n},\overrightarrow{n}-\overrightarrow
{g}_{k,i}^{\left(  r\right)  }+\overrightarrow{g}_{i,l}^{\left(  2\right)  }%
}=\alpha_{i,l}$ in the closed queueing network.

In summary, the above analysis gives%
\[
P_{\overrightarrow{n},\overrightarrow{n}^{\prime}}=\left\{
\begin{array}
[c]{lll}%
P_{\overrightarrow{n},\overrightarrow{n}-\overrightarrow{g}_{i}%
+\overrightarrow{g}_{i,l}^{\left(  1\right)  }}=p_{i,l}, & \text{if }1\leq
n_{i}\leq K, & \text{(station }\rightarrow\text{ road)}\\
P_{\overrightarrow{n},\overrightarrow{n}-\overrightarrow{g}_{k,i}^{\left(
r\right)  }+\overrightarrow{g}_{i}}=1, & \text{if }0\leq n_{i}\leq K-1, &
\text{(road }\rightarrow\text{ station)}\\
P_{\overrightarrow{n},\overrightarrow{n}-\overrightarrow{g}_{k,i}^{\left(
r\right)  }+\overrightarrow{g}_{i,l}^{\left(  2\right)  }}=\alpha_{i,l}, &
\text{if }n_{i}=K, & \text{(road }\rightarrow\text{ road, a full station)}\\
0, & \text{otherwise.} &
\end{array}
\right.
\]
At the same time, the minimal number of zero elements in the routing matrix
$\mathbf{P}$ is given by%
\[
\left[  \left(  K+1\right)  ^{N}\left(  NC+1\right)  ^{2N\left(  N-1\right)
}\right]  ^{2}-NK\left(  N-1\right)  -2N^{2}\left(  N-1\right)  CK-2N^{3}%
\left(  N-1\right)  ^{2}C^{2}%
\]
This also shows that there exist more zero elements in the routing matrix
$\mathbf{P}$.

We write a row vector%
\[
\overrightarrow{\Re}=\left(  \overrightarrow{e}\left(  \overrightarrow
{n}\right)  :\overrightarrow{e}\left(  \overrightarrow{n}\right)
\in\mathbb{E}\right)  ,
\]
where%

\[
\mathbb{E}=\left\{  \overrightarrow{e}\left(  \overrightarrow{n}\right)
:\overrightarrow{n}\in\Omega\right\}  .
\]

\begin{The}
The routing matrix $\mathbf{P}$ is irreducible and stochastic (i.e.,
$\mathbf{P1}=\mathbf{1}$, where $\mathbf{1}$ is a column vector of ones), and
there exists a unique positive solution to the following system of linear
equations%
\[
\left\{
\begin{array}
[c]{c}%
\overrightarrow{\Re}=\overrightarrow{\Re}\text{ }\mathbf{P},\\
\left(  \overrightarrow{\Re}\right)  _{1}=1,
\end{array}
\right.
\]
where $\left(  \overrightarrow{\Re}\right)  _{1}$ is the first element of the
row vector $\overrightarrow{\Re}$.
\end{The}

\textbf{Proof:} \ The outline of this proof is described as follows. It is
well-known that the routing structure of the closed queueing network indicates
that the routing matrix $\mathbf{P}$ is stochastic, and the accessibility of
each station or road of the bike sharing system shows that the routing matrix
$\mathbf{P}$ is irreducible. Thus the routing matrix $\mathbf{P}$ is not only
irreducible but also stochastic. Notice that the size of the routing matrix
$\mathbf{P}$ is $\left(  K+1\right)  ^{N}\left(  NC+1\right)  ^{2N\left(  N-1\right)  }$,
it follows from Theorem 1.1 (a) and (b) of Chapter 1 in Seneta
\cite{Sen:1981} that the left eigenvector $\overrightarrow{\Re}$ of the
irreducible stochastic matrix $\mathbf{P}$ corresponding to the maximal
eigenvalue $1$ is more than $0$, that is, $\overrightarrow{\Re}>0$, and
$\overrightarrow{\Re}$ is unique for $\left(  \overrightarrow{\Re}\right)
_{1}=1$. This completes this proof. \textbf{{\rule{0.08in}{0.08in}}}

\section{A Product-Form Solution and Performance Analysis}

In this section, we first provide a product-form solution to the steady state
joint probabilities of $N\left(  2N-1\right)  $ queue lengths in the closed
queueing network. Then we analyze performance measures of the bike sharing
system by means of the steady state joint probabilities.

Notice that%

\[
\mathbf{X}\left(  t\right)  =\left(  \mathbf{L}_{1}\left(  t\right)
,\mathbf{L}_{2}\left(  t\right)  ,\ldots,\mathbf{L}_{N-1}\left(  t\right)
,\mathbf{L}_{N}\left(  t\right)  \right)  ,
\]
where for $1\leq i\leq N$%
\begin{align*}
\mathbf{L}_{i}\left(  t\right)  = &  \left(  Q_{i}\left(  t\right)
;R_{i,1}^{\left(  1\right)  }\left(  t\right)  ,R_{i,1}^{\left(  2\right)
}\left(  t\right)  ;R_{i,2}^{\left(  1\right)  }\left(  t\right)
,R_{i,2}^{\left(  2\right)  }\left(  t\right)  ;\ldots;R_{i,i-1}^{\left(
1\right)  }\left(  t\right)  ,R_{i,i-1}^{\left(  2\right)  }\left(  t\right)
;\right.  \\
&  \left.  R_{i,i+1}^{\left(  1\right)  }\left(  t\right)  ,R_{i,i+1}^{\left(
2\right)  }\left(  t\right)  ;R_{i,i+2}^{\left(  1\right)  }\left(  t\right)
,R_{i,i+2}^{\left(  2\right)  }\left(  t\right)  ;\ldots;R_{i,N}^{\left(
1\right)  }\left(  t\right)  ,R_{i,N}^{\left(  2\right)  }\left(  t\right)
\right)  .
\end{align*}
At the same time, $\left\{  \mathbf{X}\left(  t\right)  :t\geq0\right\}  $ is
an irreducible continuous-time Markov process on state space $\Omega$ which
contains $\left(  K+1\right)  ^{N}\left(  NC+1\right)  ^{2N\left(  N-1\right)  }$
states. Therefore, the Markov process $\left\{  \mathbf{X}%
\left(  t\right)  :t\geq0\right\}  $ is irreducible and positive recurrent. In
this case, we set
\begin{align*}
\mathbf{\pi}\left(  \overrightarrow{n}\right)  = &  \lim_{t\rightarrow+\infty
}P\left\{  Q_{i}\left(  t\right)  =n_{i},1\leq i\leq N;\text{ }R_{k,l}%
^{\left(  1\right)  }\left(  t\right)  =m_{k,l}^{\left(  1\right)  }%
,R_{k,l}^{\left(  2\right)  }\left(  t\right)  =m_{k,l}^{\left(  2\right)
},\right.  \\
&  \left.  1\leq k,l\leq N\text{ with }k\neq l,\sum\limits_{i=1}^{N}n_{i}%
+\sum_{r=1,2}\sum\limits_{k=1}^{N}\sum\limits_{l\neq k}^{N}m_{k,l}^{\left(
r\right)  }=NC\right\}  .
\end{align*}

\textbf{(a) A product-form solution to the steady state joint probabilities}

The following theorem provides a product-form solution to the steady state
joint probability $\mathbf{\pi}\left(  \overrightarrow{n}\right)  $ for
$\overrightarrow{n}\in\Omega$; while its proof is easy by means of Chapter 7
in Bolch et al. \cite{Bol:2006} and is omitted here.

\begin{The}
For the closed queueing network of the bike sharing system, the steady state
joint probability $\mathbf{\pi}\left(  \overrightarrow{n}\right)  $ is given
by
\[
\mathbf{\pi}\left(  \overrightarrow{n}\right)  =\frac{1}{\mathbf{G}}%
\prod_{i=1}^{N}F\left(  n_{i}\right)  \prod_{k=1}^{N}\prod_{l\neq k}%
^{N}m_{k,l}!H^{\left(  1\right)  }\left(  m_{k,l}^{\left(  1\right)  }\right)
H^{\left(  2\right)  }\left(  m_{k,l}^{\left(  2\right)  }\right)  ,
\]
where $\overrightarrow{n}\in\Omega$, $m_{k,l}=m_{k,l}^{\left(  1\right)
}+m_{k,l}^{\left(  2\right)  }$,%
\[
F\left(  n_{i}\right)  =\left\{
\begin{array}
[c]{ll}%
\left[  \frac{e_{i}\left(  n_{i}\right)  }{\lambda_{i}}\right]  ^{n_{i}}, &
1\leq n_{i}\leq K,\\
1, & n_{i}=0,
\end{array}
\right.  \text{ }%
\]%
\[
H^{\left(  1\right)  }\left(  m_{k,l}^{\left(  1\right)  }\right)  =\left\{
\begin{array}
[c]{ll}%
\frac{1}{m_{k,l}^{\left(  1\right)  }!}\left[  \frac{e_{k,l}^{\left(
1\right)  }\left(  m_{k,l}^{\left(  1\right)  }\right)  }{m_{k,l}^{\left(
1\right)  }\mu_{k,l}}\right]  ^{m_{k,l}^{\left(  1\right)  }}, & 1\leq
m_{k,l}^{\left(  1\right)  }\leq NC,\\
1, & m_{k,l}^{\left(  1\right)  }=0,
\end{array}
\right.  ,
\]%
\[
H^{\left(  2\right)  }\left(  m_{k,l}^{\left(  2\right)  }\right)  =\left\{
\begin{array}
[c]{ll}%
\frac{1}{m_{k,l}^{\left(  2\right)  }!}\left[  \frac{e_{k,l}^{\left(
2\right)  }\left(  m_{k,l}^{\left(  2\right)  }\right)  }{m_{k,l}^{\left(
2\right)  }\xi_{k,l}}\right]  ^{m_{k,l}^{\left(  2\right)  }}, & 1\leq
m_{k,l}^{\left(  2\right)  }\leq NC,\\
1, & m_{k,l}^{\left(  2\right)  }=0,
\end{array}
\right.
\]
and $\mathbf{G}$ is a normalization constant, given by
\[
\mathbf{G}=\sum_{\overrightarrow{n}\in\Omega}\prod_{i=1}^{N}F\left(
n_{i}\right)  \prod_{k=1}^{N}\prod_{l\neq k}^{N}m_{k,l}!H^{\left(  1\right)
}\left(  m_{k,l}^{\left(  1\right)  }\right)  H^{\left(  2\right)  }\left(
m_{k,l}^{\left(  2\right)  }\right)  .
\]
\end{The}

\textbf{(b) Performance analysis}

Now, we consider three key performance measures of the bike sharing system in
terms of the steady state joint probability $\mathbf{\pi}\left(
\overrightarrow{n}\right)  $ for $\overrightarrow{n}\in\Omega$.

\textit{(1) The steady state probability of problematic stations}

In the study of bike sharing systems, it is a key task to compute the steady
state probability of problematic stations. To this end, our aim is to care for
the $i$th station with respect to its full or empty cases. Thus the steady
state probability $\Im$ of problematic stations is given by
\begin{align*}
\Im &  =P\left\{  n_{i}=0\text{ or }n_{i}=K\right\}  =P\left\{  n_{i}%
=0\right\}  +P\left\{  n_{i}=K\right\}  \\
&  =\sum\limits_{\overrightarrow{n}\in\Omega\text{ \& }n_{i}=0}\mathbf{\pi
}\left(  \overrightarrow{n}\right)  +\sum\limits_{\overrightarrow{n}\in
\Omega\text{ \& }n_{i}=K}\mathbf{\pi}\left(  \overrightarrow{n}\right)  .
\end{align*}

\textit{(2) The means of steady state queue lengths}

The steady state mean of the number of bikes parked\textit{ at }the $i$th
station is given by%
\[
\mathbf{Q}_{i}=\sum\limits_{\overrightarrow{n}\in\Omega\text{ \& }1\leq
n_{i}\leq K}n_{i}\mathbf{\pi}\left(  \overrightarrow{n}\right)  ,\text{
\ }1\leq i\leq N,
\]
and the steady state mean of the number of bikes ridden on the
$N\left(  N-1\right)  $ roads is given by%
\[
\mathbf{Q}_{0}=NC-\sum_{i=1}^{N}\left[  \sum\limits_{\overrightarrow{n}%
\in\Omega\text{ \& }1\leq n_{i}\leq K}n_{i}\mathbf{\pi}\left(  \overrightarrow
{n}\right)  \right]  ,
\]
or%
\[
\mathbf{Q}_{0}=\sum_{r=1,2}\sum_{k=1}^{N}\sum_{l\neq k}^{N}\sum
_{\overrightarrow{n}\in\Omega\text{ \& }1\leq m_{k,l}^{\left(  r\right)  }\leq
NC}^{N}m_{k,l}^{\left(  r\right)  }\mathbf{\pi}\left(  \overrightarrow
{n}\right)  .
\]

\section{Concluding Remarks}

In this paper, we provide a unified framework for analyzing closed queueing
networks in the study of bike sharing systems, and show that this framework of
closed queueing networks is interesting, difficult and challenging. We
describe and analyze a closed queueing network corresponding to a large-scale
bike sharing system, and specifically, we provide a product-form solution to
the steady state joint probabilities of $N\left(  2N-1\right)  $ queue
lengths, which leads to be able to calculate the steady state probability of
problematic stations, and more generally, to analyze performance measures of
this bike sharing system. We hope the methodology and results of this paper
can be applicable in the study of more general bike sharing systems by means
of the closed queueing networks. Along these lines, there are a number of
interesting directions for potential future research, for example:

\begin{itemize}
\item Developing effective algorithms for computing the routing matrix, the
relative arrival rates, and the steady state joint probabilities of queue lengths;

\item analyzing bike sharing systems with Markovian arrival processes (MAPs)
of customers to rent bikes, and phase type (PH) riding-bike times on the roads;

\item considering heterogeneity of bike sharing systems under an irreducible
graph with stations, roads and their connections;

\item discussing repositioning bikes by trucks in bike sharing systems with
information technologies; and

\item applying periodic MAPs, periodic PH distributions, or periodic Markov
processes to studying time-inhomogeneous bike sharing systems.
\end{itemize}

\section*{Acknowledgements}

Q.L. Li was supported by the National Natural Science Foundation of China
under grant No. 71271187 and No. 71471160, and the Fostering Plan of
Innovation Team and Leading Talent in Hebei Universities under grant No. LJRC027.

\end{document}